\documentclass[showpacs,aps,prl,twocolumn,superscriptaddress]{revtex4}
\usepackage{graphicx} 
\usepackage{dcolumn}
\usepackage{bm}
\usepackage{amssymb,amsmath}
\usepackage{epstopdf}

\begin{document}
\title{Large Magnetic Susceptibility Anisotropy of Metallic Carbon Nanotubes}

\normalsize
\author{T. A. Searles}
\affiliation{Department of Electrical and Computer Engineering, Rice University, Houston, Texas 77005, USA}

\author{Y. Imanaka}
\affiliation{National Institute for Materials Science, 3-13 Sakura, Tsukuba, Ibaraki 305-0003, Japan}

\author{T.~Takamasu}
\affiliation{National Institute for Materials Science, 3-13 Sakura, Tsukuba, Ibaraki 305-0003, Japan}

\author{H. Ajiki}
\affiliation{Photon Pioneers Center, Osaka University, Suita 565-0871, Japan}

\author{J. A. Fagan}
\affiliation{National Institute of Standards and Technology, Gaithersburg, Maryland 20899, USA}

\author{E. K. Hobbie}
\affiliation{National Institute of Standards and Technology, Gaithersburg, Maryland 20899, USA}

\author{J. Kono}
\email[]{kono@rice.edu}
\thanks{corresponding author.}
\affiliation{Department of Electrical and Computer Engineering, Rice University, Houston, Texas 77005, USA}

\date{\today}

\begin{abstract}
Through magnetic linear dichroism spectroscopy, the magnetic susceptibility anisotropy of metallic single-walled carbon nanotubes has been extracted and found to be 2-4 times greater than values for semiconducting single-walled carbon nanotubes.  This large anisotropy is consistent with our calculations and can be understood in terms of large orbital paramagnetism of electrons in metallic nanotubes arising from the Aharonov-Bohm-phase-induced gap opening in a parallel field.  We also compare our values with previous work for semiconducting nanotubes, which confirm a break from the prediction that the magnetic susceptibility anisotropy increases linearly with the diameter.
\end{abstract}


\maketitle



Unusual magnetic properties of single-walled carbon nanotubes (SWNTs) have generated much theoretical interest during the last two decades~\cite{AjikiAndo93JPSJ,AjikiAndo93BJPSJ,TianDatta94PRB,Lu95PRL,AjikiAndo95JPSJ,RocheetAl00PRB,NemecCuniberti06PRB,MarquesetAl06PRB}.
The orbital magnetic susceptibility $\chi$ of SWNTs is predicted to be large, two orders of magnitude larger than their spin magnetic susceptibility and comparable in magnitude to the well-known large diamagnetic susceptibility of graphite.  In addition, the value of $\chi$ is predicted to be strongly dependent on the strength of the applied magnetic field, the Fermi energy, and the field orientation.  Furthermore, the sign of $\chi$ can be either positive (paramagnetic) or negative (diamagnetic), depending on the nanotube chirality as well as the orientation of the magnetic field relative to the nanotube axis.  At the core of these properties is the Aharonov-Bohm effect~\cite{AjikiAndo93JPSJ,KonoRoche06CRC}, which changes the electronic band structure of the SWNTs via a tube-threading magnetic flux in a truly non-intuitive manner.

\begin{figure}
\includegraphics [scale=0.57] {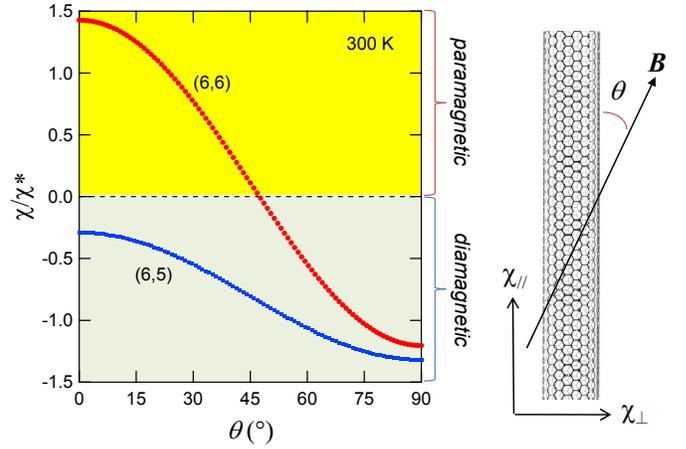}
\caption{Predicted magnetic susceptibility anisotropy of single-walled carbon nanotubes.  The magnetic susceptibility $\chi$ was calculated through a $\mathbf{k}\cdot\mathbf{p}$ model and is expressed in units of $\chi^*$ = $(2\pi\gamma/a)(\pi a^2/\phi_0)^2 a^{-2}$ = 1.46~$\times$~10$^{-4}$~emu/mol, where $\gamma$ = 6.46~eV$\cdot${\AA}, $a$ = 2.5~{\AA}, and $\phi_0 = e/h$ is the magnetic flux quantum.  The (6,6) nanotube (metallic) is paramagnetic along the tube axis and diamagnetic in the perpendicular direction ($\chi_{\parallel} > 0 > \chi_{\perp}$), whereas the (6,5) nanotube (semiconducting) is diamagnetic in all directions but most strongly diamagnetic in the perpendicular direction ($\chi_{\perp} < \chi_{\parallel} < 0$).  In both cases, $\Delta\chi = \chi_{\parallel} - \chi_{\perp} > 0$, leading to alignment in an external magnetic field $B$.}
\label{cartoon}
\end{figure}

The magnetic susceptibility anisotropy, $\Delta\chi = \chi_{\parallel} - \chi_{\perp}$, where $\chi_{\parallel}$ ($\chi_{\perp}$) is the parallel (perpendicular) susceptibility, of SWNTs has been calculated using different methods~\cite{AjikiAndo93BJPSJ,Lu95PRL,AjikiAndo95JPSJ,MarquesetAl06PRB}.  Metallic SWNTs are predicted to be paramagnetic along the tube axis and diamagnetic in the perpendicular direction ($\chi^M_{\parallel} > 0 > \chi^M_{\perp}$), whereas semiconducting SWNTs are predicted to be diamagnetic in all directions but most strongly diamagnetic in the perpendicular direction ($\chi^S_{\perp} < \chi^S_{\parallel} < 0$)~\cite{AjikiAndo93BJPSJ,Lu95PRL,AjikiAndo95JPSJ}.  See, e.g., Fig.~\ref{cartoon} for our $\mathbf{k}\cdot\mathbf{p}$ calculations for (6,6) and (6,5) nanotubes at 300~K.  Thus, all SWNTs are expected to have positive $\Delta\chi$, while metallic SWNTs are expected to have greater values of $\Delta\chi$ than semiconducting SWNTs.  A finite $\Delta\chi$ results in the alignment of SWNTs in the magnetic field direction, which, combined with the anisotropic optical properties of SWNTs, allows for an estimation of $\Delta\chi$ values for SWNTs through magneto-optical spectroscopy.  Previous magneto-optical studies found that $\Delta\chi \sim 1.5 \times 10^{-5}$~emu/mol for semiconducting SWNTs with $\sim$1~nm diameters~\cite{KonoRoche06CRC,ZaricetAl04Science,ZaricetAl04NL,IslametAl05PRB,TorrensetAl07JACS,KonoetAl07Book}, while no information on $\Delta\chi$ is currently available for metallic SWNTs.


Here, we present the first experimental estimation of the $\Delta\chi$ of metallic SWNTs through high-field magnetic linear dichroism spectroscopy of CoMoCAT nanotubes individually suspended in aqueous solution at room temperature.  The sample had a much smaller diameter distribution than the HiPco samples used in the previous studies~\cite{ZaricetAl04Science,ZaricetAl04NL}, which allowed us to clearly identify, and closely examine the magnetic field dependence of, absorption peaks for metallic nanotubes.  We made a detailed comparison of the $\Delta\chi$ of metallic and semiconducting nanotubes with similar diameters and lengths within the same sample and found that the values of $\Delta\chi$ are 2-4 times greater in metallic tubes.  We also compared our values with previous work for semiconducting nanotubes~\cite{TorrensetAl07JACS}, which confirm a break from the prediction that $\Delta\chi$ should increase linearly with the tube diameter.


Polarized magneto-optical absorption measurements on length-sorted, (6,5)-enriched CoMoCAT SWNTs were performed using the 35~T hybrid magnet at the National Institute for Materials Science in Tsukuba, Japan.  The SWNTs were suspended in 1\% sodium deoxycholate and length-sorted by dense liquid ultracentrifugation to have an average length of $\sim$500~nm. A Xe lamp, fiber-coupled to a custom optical probe, allowed for broadband white-light excitation of the $E_{11}$ metallic, $E_{22}$ semiconductor, and $E_{33}$ semiconductor interband transitions of SWNTs.  The sample was held in a cuvette with a film polarizer directly on the front face to ensure that the incident light was linearly polarized.  The transmitted light was collimated by a lens, collected with another fiber, and dispersed through a monochromator equipped with a Si CCD.  All measurements were done at $\sim$300~K.


\begin{figure}
\includegraphics [scale=0.55] {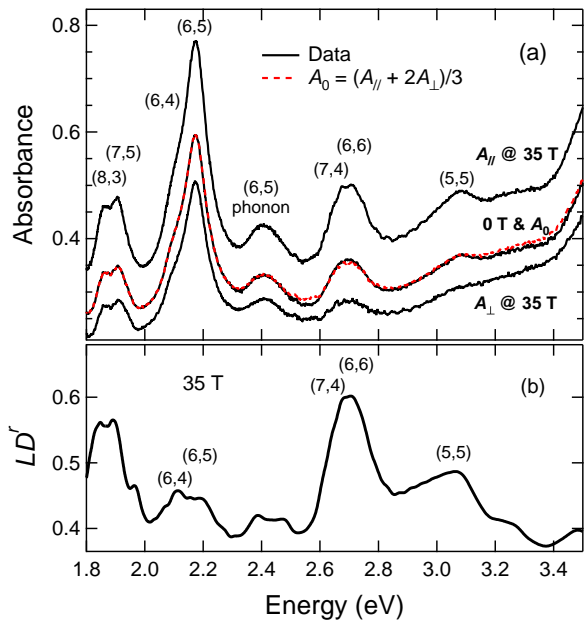}
\caption{(a) Absorption spectra (solid black) for 0~T and 35~T with all peaks assigned to specific chiralities.  No trace is intentionally offset, indicating greater (smaller) absorption for parallel (perpendicular) polarization.  The unpolarized isotropic absorbance (dashed red), calculated from the 35~T spectra via Eq.~(\ref{A0}), agrees well with the 0~T data.  (b) Reduced linear dichroism versus energy from measured data.  The largest peak is from metallic nanotubes (6,6) and (7,4).}
\label{abs}
\end{figure}
Figure~\ref{abs}(a) shows polarization-dependent optical absorption spectra at 0 and 35~T.  None of the spectra are intentionally offset, indicating an increase (decrease) in absorbance for light polarized parallel (perpendicular) to the field.  This is a direct result of magnetic alignment, together with the fact that only the light-field component parallel to the tube axis is strongly absorbed in SWNTs.  Namely, at 35~T there is a finite degree of alignment in the magnetic field direction within the nanotube ensemble, resulting in stronger (weaker) absorption for light polarized parallel (perpendicular) to the field.  From the theory of linear dichroism for an ensemble of anisotropic molecules~\cite{RodgerNordenCDLD1997}, the
following quantity can be shown to be constant, independent of the degree of alignment:
\begin{equation}
A_{0} \equiv \frac {A_{\parallel} + 2A_{\perp}}{3},
\label{A0}
\end{equation}
where $A_{\parallel}$ ($A_{\perp}$) is the absorption for light polarized parallel (perpendicular) to the orientation axis, i.e., the magnetic field direction.  We calculated $A_{0}$ at 35~T using Eq.~(\ref{A0}) and plotted it in Fig.~\ref{abs}(a) as a red dashed line.  The agreement between the calculated $A_0$ and the zero-field absorbance spectrum confirms that $A_{0}$ is independent of alignment (or $B$), because at 0~T $A_{\parallel}$ = $A_{\perp}$ = $A_0$.  We further confirmed that at any fields between 0 and 35~T the increase in $A_{\parallel}$ and decrease in $A_{\perp}$ are such that $A_{0}$ is preserved through Eq.~(\ref{A0}).

Linear dichroism, $LD = A_{\parallel}-A_{\perp}$, is a measure of the degree of alignment.  However, since it directly depends on the absorbance, $LD$ alone cannot be used for comparing different spectral features.  To adjust for differences in relative absorbances due to the fact that our sample is enriched for (6,5), the $LD$ was divided by $A_{0}$, yielding the {\em reduced} linear dichroism, $LD^{r} = LD/A_0$~\cite{RodgerNordenCDLD1997,ShaveretAl09ACS}, which is plotted in Fig.~\ref{abs}(b) for 35~T.  It is immediately evident from this plot that the spectral region where metallic peaks [(6,6) and (7,4)] exist has much larger $LD^r$ than the region where the most prominent semiconducting peaks [(6,5) and (6,4)] exist.  This is evidence that the (6,6) and (7,4) tubes are aligning more strongly than the (6,5) and (6,4) tubes.  In order to quantitatively determine the magnetic alignment properties of each chirality present in our sample, we performed detailed spectral analysis, as described below.


\begin{figure}
\includegraphics [scale=0.57] {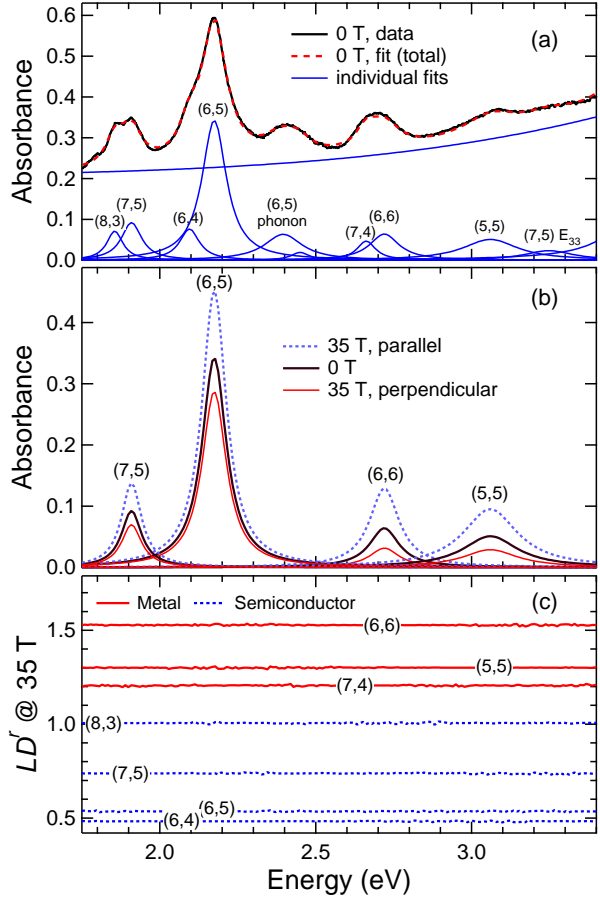}
\caption{(a) Fitting results for 0~T (b) Lorentzians for different chirality nanotubes at 0 and 35~T. (c)  $LD^{r}$vs. Energy (eV) derived from fitting for each individual chirality nanotube in our sample.  The metallic tubes (red) are higher than the semiconducting nanotubes (blue dotted).}
\label{fits}
\end{figure}

Each absorption spectrum was fit by a superposition of multiple Lorentzian peaks plus a polynomial offset.  The zero-field spectrum was fit first, as shown in Fig.~\ref{fits}(a), with each Lorentzian assigned to a specific chirality.  The offset was included to account for the contributions from the $\pi$ plasmon peaks, bundles, and light scattering~\cite{MurakamietAl05PRL,NairetAlAnalChem06,FaganAPL2007}.  For each magnetic field, the parallel and perpendicular spectra were fit simultaneously with the constraint that each Lorentzian independently satisfy Eq.~(\ref{A0}).  Figure~\ref{fits}(b) shows representative results for the (7,5), (6,5), (6,6), and (5,5) nanotubes at 0 and 35~T.  Note that we required the width and position of each Lorentzian to be the same as the values obtained at 0~T, which ensured that the calculated $LD^r$ for each chirality is constant as a function of photon energy, as shown in Fig.~\ref{fits}(c) for 35~T.  Here, each $LD^{r}$ line is a direct result of calculation from each Lorentzian of different chirality, while $LD^{r}$ in Fig.~\ref{abs}(b) was calculated from the entire spectra.  As a result, one can extract chirality-specific information on the alignment degree from Fig.~\ref{fits}(c).  It is evident that the values are much larger for the metallic (red lines) than the semiconducting (blue lines) nanotubes.

\begin{figure}
\includegraphics [scale=0.5] {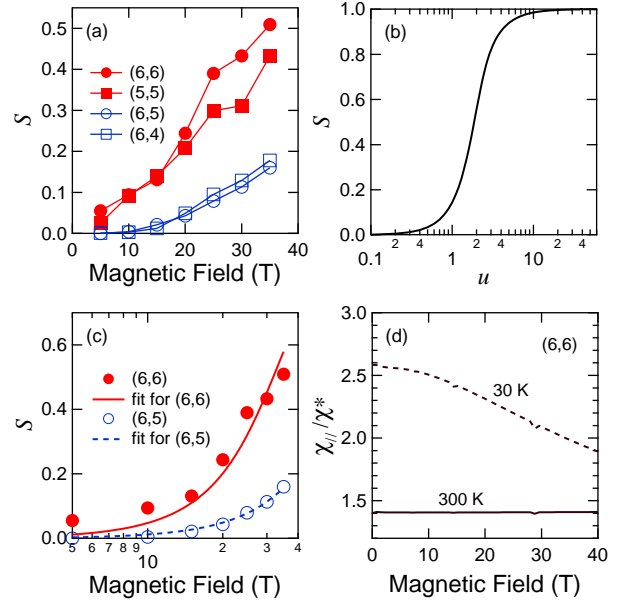}
\caption{(a) The nematic order parameter $S$ as a function of magnetic field for armchair nanotubes (5,5) and (6,6) and semiconducting nanotubes (6,5) and (6,4), calculated from the measured reduced linear dichroism values.  (b) $S$ versus $u$ [Eq.~(\ref{Su})].  (c) Comparison of $S$ versus magnetic field for (6,5) and (6,6) nanotubes.  (d) Calculated magnetic field dependence of the parallel magnetic susceptibility of (6,6) nanotube for $T$ = 30~K and 300~K in units of $\chi^*$~=~1.46~$\times$~10$^{-4}$~emu/mol.}
\label{S}
\end{figure}

For anisotropic molecules with the dipole moment fully oriented along the long axis, the reduced linear dichroism can be expressed as $LD^{r} = 3S$~\cite{RodgerNordenCDLD1997,ShaveretAl09ACS}.  Here, the nematic order parameter $S = (3\langle\cos^2\theta\rangle-1)/2$ is a dimensionless measure of the degree of alignment, being equal to zero for completely randomly oriented nanotubes and one for completely aligned nanotubes, and $\theta$ is the angle between the nanotube axis and the orientation axis (i.e., the magnetic field direction in the present case).  Values for $S$ were calculated for representative nanotubes from the $LD^r$ values in Fig.~\ref{fits}(c) and plotted as a function of magnetic field in Fig.~\ref{S}(a).  Here, we can clearly see that the ($n$,$n$)-chirality, or armchair, metallic nanotubes [(6,6) and (5,5)] show much stronger alignment than the semiconducting nanotubes [(6,5) and (6,4)].

In order to model the observed magnetic field dependence of $S$, we approximate the angular distribution of nanotubes at temperature $T$ in a magnetic field $B$ by
the Maxwell-Boltzmann distribution 
function~\cite{ZaricetAl04Science,ZaricetAl04NL,IslametAl05PRB,TorrensetAl07JACS,FujiwaraetAl99JPCB}
\begin{equation}
P_{u}(\theta) = \frac{\exp( -u^{2} \sin^{2} \theta) \sin \theta}{\int_0^{\pi/2} \exp( -u^{2} \sin^{2} \theta) \sin \theta d\theta},
\end{equation}
where $u = \sqrt \frac{B^{2}N\Delta\chi}{2k_{\rm B}T}$ is a dimensionless measure of the relative importance of the
magnetic alignment energy ($B^2 N \Delta\chi/2$) and the thermal energy ($k_{\rm B}T$), $N$ is the number of carbon
atoms in each nanotube with a length of 500~nm (in moles), and $k_{\rm B}$ represents Boltzmann's constant.
Using this distribution function and the definition of $S$, one can derive
\begin{eqnarray}
\nonumber S(u) = \int^{\pi/2}_{0} P_{u}(\theta) \left({{3 \cos^2\theta -1} \over 2}\right) d\theta \\
= \frac{-1}{2}+\left(\frac {3}{4u^{2}}\right)\left(\frac{u}{\frac{\sqrt\pi}{2}e^{-u^{2}}\mbox{erfi}(u)}-1\right),
\label{Su}
\end{eqnarray}
where $\mbox{erfi}$ is the imaginary error function [$\mbox{erfi}(u) =\frac{2}{i\sqrt{\pi}}\int _0 ^{iu} e^{-t^2} \: dt$].  Equation (\ref{Su}) is plotted in Fig.~\ref{S}(b).


We were able to fit the $S$ versus $B$ curves with Eq.~(\ref{Su}), as shown for (6,6) and (6,5) in Fig.~\ref{S}(c), to extract $\Delta\chi$ for each nanotube chirality.  However, care must be taken in using this procedure, since the values of $\Delta\chi$ are expected to depend on $B$ and $T$, especially for metallic nanotubes.  Thus, we calculated $\Delta\chi$ as a function of $B$ and $T$ for all the seven SWNTs studied in this work.  Figure~\ref{S}(d) shows the $B$ dependence of $\chi_{\parallel}$ for the (6,6) nanotube for 30~K and 300~K.  Although $\chi_{\parallel}$ decreases by $\sim$27\% at 30~K as the field increases from 0~T to 40~T, it stays constant within $\sim$0.7\% at 300~K in the same field range, ensuring that the magnetic field dependence can be neglected for our room temperature measurements.

Theoretical and experimental values of $\Delta\chi$ obtained in the present work are summarized in Table~\ref{table}.  The values for the three metallic nanotubes, (7,4), (5,5), and (6,6), are all higher than those for the semiconducting nanotubes.  In particular, the $\Delta\chi$ of armchair, or ($n$,$n$), carbon nanotubes are 2-4 times larger than those in semiconducting nanotubes, depending on the diameter.  This large difference in magnetic susceptibility anisotropy is a direct consequence of the Aharonov-Bohm physics causing the band structure to change in the form of bandgap opening in metallic nanotubes and bandgap shrinking for semiconducting nanotubes.  Specifically for metallic nanotubes, this causes a large paramagnetism in the direction along the tube axis.

\begin{table}
\caption{\label{table}Comparison of calculated and measured values of magnetic susceptibility anisotropy for seven types of SWNTs.  For each chirality ($n$,$m$), the diameter $d$, the chiral index $\nu$ = mod 3 $(n-m)$, and the chiral angle $\alpha$ are given, followed by theoretical (30~K and 300~K) and experimental (300~K) values of $\Delta\chi$.  All values for $\Delta\chi$ are $\times 10^{-5}$~emu/mol.}
\begin{ruledtabular}
\begin{tabular}{ccccccc}
$(n,m)$ & $d$ (nm) & $\nu$ & $\alpha (^\circ)$ & $\Delta\chi_{\rm th}^{\rm 30K}$ & $\Delta\chi_{\rm th}^{\rm 300K}$ & $\Delta\chi_{\rm exp}^{\rm 300K}$ \\
\hline
(6,6)& 0.83 & 0 & 30 & 5.78 & 3.92 & 3.63 \\
(5,5)& 0.69 & 0 & 30 & 4.95 & 3.39 & 3.35 \\
(7,4)& 0.77 & 0 & 21 & 5.42 & 3.70 & 2.44 \\
(8,3)& 0.78 & -1 & 15 & 1.49 & 1.46 & 2.13 \\
(7,5)& 0.83 & -1 & 24 & 1.58 & 1.55 & 1.66 \\
(6,5)& 0.76 & 1 & 27 & 1.45 & 1.42 & 1.01 \\
(6,4)& 0.69 & -1 & 23 & 1.33 & 1.29 & 1.24 \\
\end{tabular}
\end{ruledtabular}
\end{table}

Finally, the experimental values for metallic nanotubes in Table~\ref{table} do not follow a strict diameter dependence, something that is predicted for zigzag semiconducting tubes~\cite{MarquesetAl06PRB} and shown experimentally in semiconducting tubes~\cite{TorrensetAl07JACS}.  It is also important to note that the (7,4) nanotube has a larger value of $\Delta\chi$ than the semiconducting tubes, but it is not as large as those of the armchair tubes.  Unfortunately, it is the only non-armchair mod-3 nanotube in our sample, but a detailed study on a metallic-enriched sample~\cite{HarozateAl10arXiv} should yield many more metallic nanotubes to investigate in the future.


In conclusion, we have successfully measured $\Delta\chi$ for metallic single-walled carbon nanotubes for the first
time and confirmed that they are much larger than those of semiconducting nanotubes.  We also calculated the magnetic field and temperature dependence of magnetic susceptibilities of the SWNTs studied experimentally, which support the experimental findings.  Lastly, we were able to confirm previous experimental results for the chirality dependence of the magnetic susceptibility anisotropy in semiconducting nanotubes and found that this is also true for metallic nanotubes.

\begin{acknowledgments}
This work was supported by DOE-BES (through Grant No.~DEFG02-06ER46308), NSF (through Grant No.~OISE-0530220), and the Robert A.~Welch Foundation (through Grant No.~C-1509).  We thank Noe Alvarez for help with AFM length measurements and Ajit Srivastava for assistance with data analysis.  Official contribution of the National Institute of Standards and Technology; not subject to copyright in the United States.
\end{acknowledgments}


\end{document}